\documentclass[twocolumn,pre,showpacs]{revtex4}
\usepackage{graphicx}
\usepackage{amssymb}
\usepackage{array}
\usepackage{amsmath}
\usepackage{dcolumn}
\usepackage{times}
\usepackage[colorlinks = true, 
                 linkcolor = blue,
                 urlcolor  = blue,
                 citecolor = blue,
                 ]{hyperref}
\usepackage{tikz}
\usetikzlibrary{shapes,snakes}
\begin{document}

\title{Generalized fluctuation theorems for classical systems}
\author{G. S. Agarwal}
\affiliation{Department of Physics, Oklahoma State University, Stillwater, Oklahoma 74078 USA}
\email{girish.agarwal@okstate.edu}
\author{Sushanta Dattagupta}
\affiliation{Visva-Bharati, Santiniketan 731235, India}
\email{sushantad@gmail.com}
\date{\today}

\begin{abstract}
Fluctuation theorems have a very special place in the study of non equilibrium dynamics of physical systems. The form in which it is used most extensively is the Gallavoti-Cohen Fluctuation Theorem\cite{gallavoti} which is in terms of the distribution of the work $p(W)/p(-W)=\exp(\alpha W)$. We derive the general form of the fluctuation theorems for an arbitrary Gaussian Markov process and find conditions when the parameter $\alpha$ becomes a universal parameter $1/kT$. As an application we consider fluctuation theorems for classical cyclotron motion of an electron in a parabolic potential. The motion of the electron is described by four coupled Langevin equations and thus is non-trivial. The generalized theorems are equally valid for non-equilibrium steady states.
\end{abstract}
\maketitle
\section{Introduction}
Recent years have seen an upsurge of interest in generalized fluctuation dissipation like theorems for driven, nonequilibrium, dissipative systems\cite{gsaprl,baiesi,matteo,mai,speck1,ohta,evans}. Much of this renewed interest has concentrated on a new time dependent dynamical variable $W(t)$ which characterizes the work done on the system by an external, time dependent force, The natural, conjugate (to the applied force) variable is the velocity or the current. We re-examine the issue of the fluctuation theorem (FT) when the velocity and force fields are multidimensional. Typically the stochastic dynamics of many systems can be adequately described in terms of a Gaussian Markov process as for example is the case of the Brownian motion. After focusing on some general results we present specific, illustrative expressions for the much studied problem of dissipative cyclotron motion, which is necessarily two-dimensional in the plane normal to the applied magnetic field. Our aim is to critically assess the validity of the oft quoted fluctuation theorems when the two dimensional cyclotron motion is both symmetric and asymmetric. 

This FT has been subject to much recent scrutiny, especially in the context of more than one-dimension\cite{hurt,soodprl,soodpre}. The question asked is: Are fluctuations symmetric? For instance, in the case of non-equilibrium fluctuations in two-dimensional systems, does the GCFT relation apply for forces applied in $x$ and $y$-directions? With regard to this issue, new fluctuation theorems going beyond GCFT, have been proposed that deal with so-called `isometric' fluctuations. Recent experiments on a tapered rod undergoing Brownian motion in a sea of spherical beads have seen evidence for such isometric fluctuations, due apparently to the inherent anisotropy of the system\cite{soodpre}.

The organization of this paper is as follows-In Sec II we introduce the FT for a general N-degree of freedom, classical dissipative system, in terms of the work done on the system by an applied time dependent force. We prove that the distribution of work follows a Gaussian as long as the underlying stochastic process for the systems dynamics is Gaussian and Markovian. In Sec III we derive conditions for the determination of the parameters that characterize the Gaussian distribution of the work. We also seek conditions under which the parameter $\alpha$ in the FT $p[W]/p[-W]=\exp[\alpha W]$ has a universal form $1/KT$. In Sec IV we illustrate our results of Secs II and III for the case of the dissipative cyclotron motion in a confined parabolic potential. We present explicit forms for $\langle W \rangle$ and fluctuation in $W$ in terms of the temperature $T$ and the cyclotron frequency $\omega_c$. The results of Sec III, though illustrated in Sec IV for the symmetric two dimensional cyclotron motion, have much more general validity as discussed in the concluding Sec V.  
\section{The Fluctuation Theorem for a system with N degrees of freedom}
In this section we give a derivation of the FT for a system with N degrees of freedom. We will assume that the system's dynamics is described by a Gaussian Markov process. Let us denote the coordinates by $x_1, x_2, x_3, ......x_N$ and kinematic momenta, scaled by mass, by $v_1, v_2, v_3,....v_N$. The Langevin equations describing system's dynamics are assumed to have the form 
\begin{eqnarray}
\dot{x}_i &=& v_i \,, \label{dyn-eq1}\\
\dot{v}_i &=& \sum_j A_{ij}x_j + \sum_j B_{ij} v_j + f_i (t)+ \eta_i \,. \label{dyn-eq2}
\end{eqnarray}
Here $A_{ij}$ and $B_{ij}$ contain both dissipative as well as the coherent aspects of the dynamics. The $\eta_i$'s are the delta correlated Gaussian random process with zero mean. The $f_i (t)$ is the external force that acts on the variable $v_i$. As an example a force on the $x_1$ degree of freedom would appear as a driving term in the equation for the momentum $\dot{v}_1$. The work done by the external force on the system is 
\begin{eqnarray}
\label{work-force}
W &=& \sum_{j=1}^N \int_0 ^t d\tau f_j (\tau)\dot{x}_j (\tau)m_j \nonumber\\
&=& \sum_{j=1}^N \int_0 ^t d\tau f_j (\tau) v_j (\tau) m_j \,.
\end{eqnarray}
\subsection{Distribution of work}
We first show that the distribution of $W$ is Gaussian. To do this we construct the characteristic function of $W$
\begin{eqnarray}
C(h) &=& \langle e^{ihW} \rangle \nonumber\\
&=& \langle e^{ih \sum_i m_i \int_0 ^t d\tau f_i (\tau) \dot{x}_i (\tau)} \rangle \nonumber\\
&=& e^{ih \langle W \rangle} \langle e^{ih \sum_i m_i \int_0 ^t d\tau f_i (\tau) (v_i (\tau) - \langle v_i (\tau)\rangle)} \rangle \,,
\label{average-added}
\end{eqnarray}
where
\begin{equation}
\langle W \rangle = \sum m_i \int_0 ^t d\tau f_i (\tau) \langle v_i (\tau) \rangle \,.
\end{equation}
It should be borne in mind that $\langle v_i (\tau) \rangle$ depends linearly on the external forces. The average appearing in Eq.(\ref{average-added}) is just the functional for Gaussian process which is well known. Using this Eq.(\ref{average-added}) becomes
\begin{equation}
C(h) = e^{ih \langle W \rangle} e^{-\frac{1}{2}h^2 \sigma^2}\,,
\end{equation}
where
\begin{eqnarray}
\sigma^2 &=& \sum \sum \int_0 ^t d\tau_1 \int_0 ^t d\tau_2 f_i (\tau_1)f_j (\tau_2)m_i m_j \nonumber\\
&\times& (\langle v_i (\tau_1)v_j (\tau_2)\rangle  - \langle v_i (\tau_1)\rangle \langle v_j (\tau_2)\rangle) \,.
\end{eqnarray}
Hence $p(W)$, which is the Fourier transform of $C(h)$, is Gaussian:
\begin{equation}
p(W)=\frac{1}{\sqrt{2\pi}\sigma}\exp\{-\frac{(W-\langle W \rangle)^2}{2\sigma^2}\} \,.
\end{equation}
Therefore,
\begin{equation}
\frac{p(-W)}{p(W)}=\exp(-\alpha W)\,,
\label{imp1}
\end{equation}
where
\begin{equation}
\label{zeta}
\alpha = 2\langle W \rangle / \sigma^2 \,.
\end{equation}
The Eq.(\ref{imp1}) is the generalized Fluctuation Theorem valid for any Gaussian Markov process. The parameter $\alpha$ can in principle depend on $f(t)$. In many cases discussed in literature the parameter $\alpha$ is a universal constant $1/KT$ where $K$ is the Boltzmann constant and $T$ the temperature. We then recover the FT of Gallavoti and Cohen. We will see next that in order to obtain the universal constant, the system's dynamics has to satisfy certain specific conditions.
\subsection{Determination of $\alpha$ using dynamical equations (\ref{dyn-eq1}) and (\ref{dyn-eq2})}
We write Eqs.(\ref{dyn-eq1}) and (\ref{dyn-eq2}) using matrices. We define column marices $\psi$, $N$, $F$ with $2N$ components arranged as $\psi_i = x_i,$; $\psi_{i+N} = v_i,$; $N_{i+N} = \eta_i ,$, $F_{i+N} = f_i,$ otherwise $N_i$ and $F_i$ are zero. Then we write Eqs.(\ref{dyn-eq1}) and (\ref{dyn-eq2}) as
\begin{equation}
\dot{\psi} = M\psi + N + F,.
\label{sum-eq}
\end{equation}
The matrix $M$ is easily constructed in terms of $A$ and $B$ matrices. Then 
\begin{eqnarray}
\psi(t) &=& V(t)\psi(0) + \int_0 ^t V(t-\tau) F(\tau) d\tau \nonumber\\
&+& \int_0 ^t V(t-\tau) N(\tau) d\tau\,,~~~V(t)=e^{Mt} \,,
\end{eqnarray}
and hence
\begin{equation}
\langle \psi(t) \rangle  = \int_0 ^t V(t-\tau) F(\tau) d\tau \,.
\end{equation}
Using Eq.(\ref{sum-eq}) in Eq.(\ref{work-force}) we get the mean value of the work
\begin{eqnarray}
\label{mean-value-work}
\langle W \rangle &=& \sum_i \sum_j \int_0 ^t d\tau_1 \int_0 ^{\tau_{1}} d\tau_2 f_i (\tau_1) \nonumber\\
&\times& v_{i+N,j+N} (\tau_1 - \tau_2) f_j (\tau_2)m_i \,.
\end{eqnarray}
Using Eq.(\ref{work-force}) we obtain the fluctuation in $W$ as
\begin{eqnarray}
\sigma^2  &=& \sum_i \sum_j \int_0 ^t \int_0 ^t d\tau_1 d\tau_2 m_i m_j f_i (\tau_1)f_j (\tau_2) \nonumber\\ &\times& \langle v_i  (\tau_1)v_j (\tau_2) \rangle \,,
\label{sigma-square}
\end{eqnarray}
where the average in Eq.(\ref{sigma-square}) is to be calculated using the Langevin equations (\ref{sum-eq}) with $F=0$. It can be shown that
\begin{equation}
\langle v_i  (\tau_1)v_j (\tau_2) \rangle = \sum_{l=1}^{2N} [V(\tau_1 - \tau_2)]_{i+N, l} \langle \psi_l v_j \rangle \,,
\end{equation}
and hence (\ref{sigma-square}) reduces to 
\begin{eqnarray}
\label{new-sigma-square}
\sigma^2  &=& \sum_i \sum_j \int_0 ^t \int_0 ^t d\tau_1 d\tau_2 m_i m_j f_i (\tau_1)f_j (\tau_2) \nonumber\\
&\times& \sum_{l=1}^{2N} [V(\tau_1 - \tau_2)]_{i+N, l} \langle \psi_l v_j \rangle \,.
\end{eqnarray}
The parameter $\alpha$ is obtained by substituting (\ref{mean-value-work}) and (\ref{new-sigma-square}) in Eq.(\ref{zeta}). We have the most general form now for the work fluctuation and its mean value in terms of the dynamics of the system. The Eqs.(\ref{imp1}), (\ref{mean-value-work}) and (\ref{new-sigma-square}) are the key results of this paper.
\section{Conditions for $\alpha$ to be universal constant}
The parameter $\alpha$ as determined by Eqs.(\ref{mean-value-work}) and (\ref{new-sigma-square}) can depend on the details of the system and the forces. Clearly we need to find conditions when $\sigma^2$ is proportional to $\langle W \rangle$. From Eq.(\ref{new-sigma-square}) it is clear that we need at least the steady state fluctuations such that
\begin{equation}
\langle \psi_l v_j \rangle  = D_j \delta_{lj+N}\,,~~ D_j  = \langle v_j ^2 \rangle \,,
\label{new-average-psi}
\end{equation}
then $\sigma^2$ reduces to 
\begin{eqnarray}
\sigma^2  &=& \sum_i \sum_j \int_0 ^t \int_0 ^t d\tau_1 d\tau_2 f_i (\tau_1)f_j (\tau_2) \nonumber\\ &\times& v_{i+N, j+N} (\tau_1 - \tau_2) D_j m_i m_j \,.
\label{sigma-square-dj-form}
\end{eqnarray}
If the force acts only on one variable say, $i=\kappa$ i.e. $f_i = \delta_{i\kappa}f$, then (\ref{mean-value-work}) and (\ref{sigma-square-dj-form}) reduce to
\begin{equation}
\sigma^2 =2mD_{\kappa} \langle W \rangle \,,
\end{equation}
\begin{eqnarray}
2\langle W \rangle  &=& \int_0 ^t \int_0 ^{t} \sum_{ij} d\tau_1 d\tau_2 f_{\kappa} (\tau_1)f_{\kappa} (\tau_2)\nonumber\\ &\times& v_{\kappa + N, \kappa+N} (\tau_1 - \tau_2) m \,.
\label{work-average}
\end{eqnarray}
In this case $\alpha$ reduces to
\begin{equation}
\alpha = 2\langle W \rangle  / 2\langle W \rangle D_{\kappa} = 1/mD_{\kappa} \,.
\label{short-zeta}
\end{equation}
The equipartition theorem gives $mD_{\kappa} = KT$ and hence 
\begin{equation}
\alpha = 1/KT\,.
\end{equation}
Another possibility is when $m_j D_j$ is independent of the index $j$, then $\alpha$ reduces to the universal constant. Thus the conditions when $\alpha$ reduces to a universal constant for a multidimensional system are (I) Eq.(\ref{new-average-psi}) and the force acts only on one variable or (II) $m_j D_j$ must be independent of the index $j$. For a system with one degree of freedom like a harmonic oscillator, like that obtained by linearizing the non-linear potential around some mean value, we have a steady state described by $\exp\lbrace -\left(\frac{p^2}{2m} +\frac{1}{2}m\omega^2 x^2 \right)/KT \rbrace$. Then the steady state mean value is such that $\langle xp \rangle = 0$. The above conditions are satisfied and the FT with $\alpha = 1/KT$ holds almost trivially.
\section{Fluctuation theorem for dissipative classical cyclotron motion}
We will now discuss the applicability of the general results of Sec.II to the case of an electron which is in a magnetic field $\vec{B}$ applied along the $z$ axis. The electron is further confined in a parabolic potential\cite{sdgprl}. The confining potential is especially important in the quantum dynamics. The Hamiltonian is given by 
\begin{equation}
H = \frac{1}{2m} \left( \vec{p}-\frac{e}{c}\vec{A} \right)^2 + \frac{1}{2}m\Omega^2 (x^2 + y^2),
\end{equation}
with $A_x = \frac{1}{2}By,~~A_y = -\frac{1}{2}Bx$. Here $\Omega$ is the frequency of the trap potential. The Langevin equations for this system have been derived earlier and are given by\cite{sdgbook}
\begin{equation}
\dot{x}_j = v_j\,, v_j = \left( p_j - \frac{e}{c}A_j \right)/m \,,
\end{equation}
and
\begin{equation}
\dot{v}_j = \eta_j - \gamma v_j - \omega_c \sum_k v_k S_{kj} - \Omega^2 x_j \,,
\label{langevin-cyclotron}
\end{equation}
where $j=x,y$. Here $\gamma$ is the damping coefficient, $\omega_c$ is the cyclotron frequency, $S = \left( \begin{array}{cc} 0 & 1 \\ -1 & 0 \end{array} \right)$. The $\eta_j$'s are Gaussian processess with 
\begin{eqnarray}
\langle \eta_j (t) \rangle &=& 0 \\
\langle \eta_j (t) \eta_{j^{\prime}} (t^{\prime}) \rangle &=& \delta_{jj^{\prime}} \delta (t-t^{\prime})2m\gamma KT\,.
\end{eqnarray}
In the steady state the distribution is proportional to 
\begin{equation}
\exp\bigg\{-(\frac{1}{2}m[v_x ^2 + v_y ^2]+\frac{1}{2}m\Omega^2 [x^2 + y^2])/KT \bigg\}
\end{equation}
and hence
\begin{equation}
\langle v_x ^2 \rangle  = \langle v_y ^2 \rangle = \Omega^2 \langle x^2 \rangle = \Omega^2 \langle y^2 \rangle = KT/m \,.
\end{equation}
All terms like $\langle v_x v_y \rangle$ etc are zero. Let us now consider a force on the electron say by a time dependent electric field in $x$ direction. Thus the work variable is
\begin{equation}
W= m\int_0 ^t d\tau f(\tau) v_x (\tau)\,.
\end{equation}
The Eqs. (\ref{mean-value-work}) and (\ref{new-sigma-square}) for the present problem reduce to $(N=2)$
\begin{equation}
\langle W \rangle  = m\int_0 ^t d\tau_1 \int_0 ^{\tau_1}d\tau_2 f(\tau_1)f(\tau_2)V_{33}(\tau_1 - \tau_2)\,,
\end{equation}
and
\begin{eqnarray}
\sigma^2 &=& m^2 \int_0 ^t \int_0 ^t  d\tau_1  d\tau_2 f(\tau_1)f(\tau_2) \sum_{l=1}^{2N} [V(\tau_1 - \tau_2)]_{3,l} \langle \psi_l v_1 \rangle \nonumber\\
&=& m^2 \int_0 ^t \int_0 ^t  d\tau_1  d\tau_2 f(\tau_1)f(\tau_2) V_{33}(\tau_1 - \tau_2) \langle v_1 ^2 \rangle \,.
\end{eqnarray}
Thus
\begin{equation}
\sigma^2  = 2m \langle v_1 ^2 \rangle . \langle W \rangle  = 2KT \langle W \rangle
\end{equation}
which on substituting in (\ref{zeta}) yields the universal value of $\alpha  = 1/KT$. We have thus shown that the FT in the standard form is valid for the dissipative cyclotron motion though it is described by a four dimensional Gaussian Markov process.

The matrix element $V_{33} (\tau)$ is obtained from the solution of (\ref{langevin-cyclotron}) for $v_x (\tau)$ under the initial conditions $x_1 (0)= x_2 (0) =0$, $v_y (0)=0$, $v_x (0)=1$, $\eta \rightarrow 0$. The Laplace transform of $V_{33} (\tau)$ is found to be
\begin{equation}
\hat{V}_{33}(z) = \left( z+\gamma+\frac{\Omega^2}{z}+\frac{\omega_c ^2}{z+\gamma+\frac{\Omega^2}{z}}\right)^{-1} \,.
\end{equation}
For $\Omega \rightarrow 0$, $\hat{V}_{33}(z)$ yields
\begin{equation}
V_{33}(\tau) = e^{-\gamma \tau}\cos \omega_c \tau
\end{equation}
Given the applied force's functional form, one can compute the explicit results for $\langle W \rangle$ and $\langle W^2 \rangle$, as shown below.

For simplicity we present below the explicit results for $\Omega=0$. Neglecting transient terms (proportional to the initial velocities), the solution of Eq.(\ref{langevin-cyclotron}) for $v_x (t)$ reads:
\begin{eqnarray}
v_x (t) &=& \frac{e^{-\gamma t}}{m} \int_0 ^t d\tau e^{\gamma \tau} \bigg\{[\eta_x (\tau)\cos\omega_c (t-\tau) \nonumber\\
&-& \eta_y (\tau) \sin\omega_c (t-\tau)]\nonumber\\ 
&+& f(t)\cos\omega_c (t-\tau)\bigg\}\,.
\end{eqnarray}
Thus,
\begin{equation}
\langle v_x (t) \rangle = \frac{e^{-\gamma t}}{m} \int_0 ^t d\tau e^{\gamma \tau} \cos\omega_c (t-\tau)\,,
\end{equation}
and therefore,
\begin{eqnarray}
\langle W(t) \rangle  &=& \frac{1}{m} \int_0 ^t dt_1 \int_0 ^{t_1} dt_2 f(t_1)f(t_2) e^{-\gamma (t_1 - t_2)}\nonumber\\
&\times& \cos\omega_c (t_1  -t_2)\,.
\label{w-average-classical}
\end{eqnarray}

Our next query is with regard to $\langle W^2 (t) \rangle$ which is given by
\begin{equation}
\langle W^2 (t) \rangle  = \int_0 ^t dt_1 \int_0 ^t dt_2 f(t_1)f(t_2) \langle v_x (t_1)v_x (t_2) \rangle \,,
\end{equation}
which, for the sake of convenient time ordering $(t_1 > t_2)$, can be rewritten as:
\begin{equation}
\langle W^2 (t) \rangle  = 2\int_0 ^t dt_1 \int_0 ^{t_1} dt_2 f(t_1)f(t_2) \langle v_x (t_1)v_x (t_2)\,,
\end{equation}
where the velocity-velocity correlation is to be obtained from Eq.(\ref{langevin-cyclotron}). In the steady state ($t_1, t_2 \rightarrow \infty$) with $t_1 - t_2 =$ finite, we find
\begin{eqnarray}
\langle W^2 (t) \rangle  &=& \frac{2KT}{m}\int_0 ^t dt_1 \int_0 ^{t_1} dt_2 f(t_1)f(t_2) e^{-\gamma (t_1 - t_2)}\nonumber\\
&\times& \cos\omega_c (t_1 - t_2)\,.
\label{wsquare-average-classical}
\end{eqnarray}
Comparing with Eq.(\ref{w-average-classical}), we find
\begin{equation}
\langle W^2 (t) \rangle  = 2KT \langle W(t) \rangle \,,
\end{equation}
and hence, the parameter $\alpha$ in Eq.(\ref{zeta}) is identical to:
\begin{equation}
\alpha = (KT)^{-1}\,.
\end{equation}
Thus,
\begin{equation}
p(-W)/p(W) = \exp(-W/kT)\,,
\end{equation}
validating the GCFT, in the present case.
\section{Conclusions}
We have derived a class of generalized fluctuation theorems valid for a system described by a Gaussian Markov process. The derivation is fairly broad-based and thus this class of FT’s would also be valid for non-equilibrium steady states and not just for thermal equilibrium. We then present conditions on the Markovian dynamics so that we recover standard form of the FT. We apply the results of Secs II and III to dissipative cyclotron motion and show the applicability of the FT even though the dissipative dynamics is in a four dimensional phase space. Our general results would be required, in the case of the so-called ``isometric" fluctuations, if the trapping potential is anisotropic and the friction coefficients in $x$ and $y$ directions are non-identical\cite{soodprl,soodpre,hurt}.

To this date much of the applications of the FT  have been to classical, non-equilibrium and driven systems in the realm of complex fluids and biological processes\cite{speck2,seifert,chau,zon,seif}. In as much as many of these systems are governed by Brownian and
generalized Brownian dynamics, the present results for generalized Gaussian Markov processes are expected to have useful relevance.

In a subsequent paper, we shall consider the extension of the present study to a hitherto less-explored area of quantum fluctuations, with specific focus on the dissipative cyclotron motion. Most of the classical results derived for the latter case would emerge as natural, special expressions in the appropriate classical limit. 

We thank Ajay Sood for useful discussions. GSA thanks M. Barma, Director, TIFR for hospitality during the course of this work.


\end{document}